\documentstyle[12pt]{article}

\def\dspace{\baselineskip = .30in}

\def\VEV#1{\left\langle #1\right\rangle}
\def\Tilde#1{\widetilde {#1}}

\begin{document}
\title{Spontaneous CP violation in the Supersymmetric Higgs
Sector\thanks{Supported in part by Department of Energy Grant
\#DE-FG02-91ER406267}}

\author{{\bf K.S. Babu} and {\bf S.M. Barr}\\ Bartol Research Institute\\
University of Delaware\\ Newark, DE 19716}

\date {}
\maketitle
\begin{abstract}

Spontaneous CP--violation in the minimal supersymmetric standard model
with a gauge singlet and a cubic superpotential is examined.  Although
the tree--level Higgs potential conserves CP, it is shown that with the
inclusion of the one--loop top--quark radiative effects, CP may be
broken spontaneously.  The CP--violating minimum requires two neutral
$(h_1,h_2)$ and one charged $(H^\pm)$ Higgs--bosons to be relatively
light with $m_{h_1}+m_{h_2} \stackrel{_<}{_\sim} 100~GeV$ and
$m_{H^\pm} \stackrel{_<}{_\sim} 110~GeV$.
The electric dipole moment of the electron is in the observable range of
$({1 \over 3}~{\rm to}~3) \times 10^{-27}$e-cm.

\end{abstract}
\newpage
\dspace

There are a number of reasons why it is interesting to study CP
violation arising from the Higgs sector.  Such CP violation could
lead to large and soon--observable electric dipole moments for the
electron and the neutron [1].  It may have a role to play in
baryogenesis in the early universe [2].  It is even possible that
this may be the origin of the CP violation observed in the Kaon
system [3].
And such Higgs sector effects may be readily studied at present
and future colliders.

In this Letter we explore CP violation  phenomenology in the Higgs
sector of low--energy supersymmetry, in the simplest model where such
effects can appear [4].
That model is the minimal supersymmetric standard
model (MSSM) to which a gauge--singlet Higgs superfield, ${N}$, has been
added with the requirement that the superpotential
contain only cubic terms [5].
There is no significant CP violation in the parameters of the Higgs
effective potential (see below),
so that any CP violation in the Higgs sector of the
model must be spontaneous.

One of the main results we report is that spontaneous CP
violation is indeed possible in this cubic singlet model.  It has been
shown by Ramao [6] that at tree--level
spontaneous CP violation does not occur in this
model.  What allows
spontaneous CP violation to occur in our case, in spite of Ramao's
theorem, is the radiative contribution to the effective potential due to
a heavy top quark [7].  A striking feature of this model is that if CP
is spontaneously broken then two of the neutral Higgs must be quite
light.  This is actually connected to Ramao's theorem [6]: as the top--quark
radiative effects are ``turned off'' the mass--squared of one (or more) of
these light Higgs goes negative, so that there remains no
CP--violating minimum.
We show that at best the masses of the
two lightest neutral scalars add up to not much more than $M_Z$.
In addition, the mass of the
charged Higgs is found to be less than about $110$ GeV.  In this scenario the
prospects for detecting these Higgs particles at LEP and LEP-200 are
very good indeed.  Moreover, since two of the neutral Higgs fields are light,
the electron electric dipole moment (EDM) comes out in the
experimentally accessible range of $(10^{-28}~{\rm to}~3 \times 10^{-27})$
e-cm (typically).

Only the radiative effects of the top--quark will be significant, so
that the relevant terms in the superpotential are
\begin{equation}
W=\lambda {H}_1{H}_2{N} + {1 \over 3}k {N}^3+h_t
{Q}{H}_2{T}^c~~.
\end{equation}
Here ${H}_1 = \left(\matrix{{H}_1^{0 *} \cr -{H}_1^-}
\right),~~{H}_2=\left(\matrix{{H}_2^+ \cr {H}_2^0}\right)$,
so that ${H}_1{H}_2 = \epsilon^{ab}{H}_1^a{H}_2^b =
({H}_1^{0 *} {H}_2^0+{H}_1^-{H}_2^+)$.
The scalar potential for the fields $H_1,H_2$ and $N$ is then given by
$V = V_0 + V_{\rm top}$, where
\begin{eqnarray}
V_0 &=& {1 \over 8} (g_1^2+g_2^2)(|H_1|^2-|H_2|^2)^2+{1 \over 2} g_2^2(
|H_1|^2|H_2|^2-|H_1H_2|^2)+ \nonumber \\
&~& \lambda^2\left(|H_1H_2|^2+|N|^2(|H_1|^2+|H_2|^2)
\right) +
k^2|N|^4+\lambda k(H_1H_2N^{*2}+H.c.) + \nonumber \\
&~& \lambda A_\lambda(H_1
H_2N+H.c.)+{1 \over 3}kA_k(N^3+H.c.) + m_1^2|H_1|^2+m_2^2|H_2|^2+
m_3^2|N|^2 \nonumber \\ \\
V_{\rm top} &=& {3 \over {16 \pi^2}}[(h_t^2|H_2|^2+M_{sq}^2)^2
{\rm ln}\{(h_t^2|H_2|^2+M_{sq}^2)/Q^2\} \nonumber \\
&~& -\{h_t^2|H_2|^2{\rm ln}(h_t^2|H_2|^2/Q^2)\}]~~.
\end{eqnarray}
Here $|H_1|^2 \equiv H_1^{\dagger}H_1$ etc.  $V_{\rm top}$ denotes the
leading--log top--quark one--loop radiative correction.  We have
assumed degenerate squarks, $M_{\Tilde{t}_L}^2 = M_{\Tilde{t}_R}^2 =
M_{sq}^2 \gg (174~GeV)^2$.

It is easy to see that all the parameters in eq. (2) can be made real by
field redefinitions, except the ratio $r \equiv A_\lambda/A_k$.  We
assume that parameter, too, is real.  (It would be approximately so
in most realistic scenarios [8].)
To examine the CP--violating minimum, let us define
\begin{eqnarray}
D & =& \lambda k \nonumber \\
E &=& \lambda A_\lambda /|N| = r \lambda A /|N| \nonumber \\
F &=& {1 \over 3} k A |N|/|H_1^0H_2^0|~~.
\end{eqnarray}
Here and throughout $A \equiv A_k$.  Let us further define the phases of
the fields $H_1^0, H_2^0$ and $N$ as follows:
\begin{eqnarray}
{\rm arg}(H_1^*H_2N) = \eta \nonumber \\
{\rm arg}(N^3) = \zeta ~~.
\end{eqnarray}
There are three terms in eq. (2) that depend on $\eta$ and $\zeta$.
Extremizing the potential with respect to these phases gives the
CP--violating solution to be
\begin{eqnarray}
{\rm cos}\eta &=& {1 \over 2}\left({{DF}\over {E^2}}-{D \over F} -{F
\over D}\right) \nonumber \\
{\rm cos}\zeta &=& {1 \over 2} \left({{DE}\over {F^2}}-{D \over E}
-{E \over D}\right) \nonumber \\
{\rm cos}(\eta - \zeta) &=& {1 \over 2}\left({{EF}\over {D^2}} -{E \over F}
-{F \over E}\right)  \\
E{\rm sin}\eta &=& -F{\rm sin}\zeta
\end{eqnarray}
These equations can be represented graphically by the triangle in
fig. 1.  Equations (6) and (7) are respectively the law of cosines and
the law of sines.

Denoting the expectation values (VEVs)
of the neutral fields by $\VEV{H_1} = v_1,~
\VEV{H_2} = v_2$ and $\VEV{N} = x$, one can expand the fields around
their minimum.
\begin{eqnarray}
H_1 &=& v_1 + {1 \over {\sqrt{2}}} {{v_1} \over {|v_1|}}\left[\Phi_1-i{{|v_2|}
\over v} \Phi_3\right] \nonumber \\
H_2 &=& v_2 + {1 \over {\sqrt{2}}} {{v_2} \over {|v_2|}}\left[\Phi_2+i{{|v_1|}
\over v} \Phi_3\right] \nonumber \\
N &=& x + {1 \over {\sqrt{2}}} {x \over {|x|}}\left[\Phi_4+i
\Phi_5\right]
\end{eqnarray}
with $v^2 = |v_1|^2+|v_2|^2$.  In the absence of CP--violation,
$\Phi_{1,2,4}$ would be scalars and $\Phi_{3,5}$ pseudoscalars.  One
pseudoscalar gets eaten by the $Z$.  (We are working in the unitary
gauge.)

It is straightforward to expand the potential in eq. (2)-(3) about the
CP--violating extremum to obtain the mass-squared matrix of $\Phi_i$,
$i=1,2..,5$.  In doing this it is convenient to adopt a common approach
of choosing $Q^2$ in eq. (3) so that at the minimum of $V_0$ the
relation $V^{\prime}_{\rm top} = 0$ is satisfied.  (Then $V_{\rm top}$
does not contribute explicitly to the equations for $v_1,v_2$ and
$x$, it only corrects the (22) entry of the
Higgs mass--squared matrix.)
The mass--squared matrix elements are
\begin{eqnarray}
{\cal M}_{11}^2 &=& \lambda_1 v_1^2 +3 r \lambda^2 v_2^2 \nonumber \\
{\cal M}_{22}^2 &=& (1+\Delta)\lambda_1 v_2^2 + 3 r \lambda^2v_1^2
\nonumber \\
{\cal M}_{33}^2 &=& 3 r \lambda^2 v^2 \nonumber \\
{\cal M}_{12}^2 &=& -\lambda_1 v_1 v_2 +(2-3r)\lambda^2v_1 v_2
\nonumber \\
{\cal M}_{13}^2 &=& {\cal M}_{23}^2 = 0 \nonumber \\
{\cal M}_{44}^2 &=& r(4r-1)A^2 + 36 r^2 \lambda^2v_1^2v_2^2/x^2+
4r(6r-1)\lambda A (v_1v_2/x){\rm cos}\eta \nonumber \\
{\cal M}_{55}^2 &=& 3 r A^2 + 12 r \lambda^2 v_1^2v_2^2/x^2+
12 r \lambda A (v_1v_2/x){\rm cos}\eta \nonumber \\
{\cal M}_{45}^2 &=& 4 r \lambda A (v_1v_2/x){\rm sin}\eta\nonumber \\
{\cal M}_{14}^2 &=& -r \lambda A v_2 {\rm cos}\eta + 2 \lambda^2 x v_1
- 6 r \lambda^2 v_2^2 v_1/x \nonumber \\
{\cal M}_{24}^2 &=& -r \lambda A v_1 {\rm cos}\eta + 2 \lambda^2 x v_2
- 6 r \lambda^2 v_1^2 v_2/x \nonumber \\
{\cal M}_{34}^2 &=& r \lambda A v {\rm sin}\eta \nonumber \\
{\cal M}_{15}^2 &=& -3 r \lambda A v_2 {\rm sin}\eta \nonumber \\
{\cal M}_{25}^2 &=& -3 r \lambda A v_1 {\rm sin}\eta \nonumber \\
{\cal M}_{35}^2 &=& -3 r \lambda A v {\rm cos}\eta - 6 r \lambda^2 v_1v_2
v/x
\end{eqnarray}
where $\lambda_1 = {1 \over 2}(g_1^2+g_2^2) = M_Z^2/v^2$, $A \equiv A_k$
and $rA \equiv A_\lambda$.  Here and henceforth $v_1,v_2$ and $x$ stand
for the absolute values of the VEVs.  $\Delta$ in the second line of
eq. (9) is the top--quark radiative correction defined to be
\begin{equation}
\Delta = {1 \over \lambda_1} {{3 h_t^4} \over {4 \pi^2}}
\left({\rm ln}\left({{M_{sq}^2}\over {m_t^2}}\right)+p\right)~~.
\end{equation}
The parameter $p$ represents possible non--logarithmic corrections which
may be significant if $\Tilde{t}_L-\Tilde{t}_R$ mixing is not
negligible [9].
The physical charged Higgs boson mass is given by
\begin{equation}
m_{H^{\pm}}^2 = M_W^2+(3r-1)\lambda^2 v^2~~.
\end{equation}
Since $\lambda_1$ and $v= \sqrt{v_1^2
+v_2^2} = 174~GeV$ are known, our parameter
set is $\{{\rm tan}\beta, x, A, r,\lambda, {\rm cos}\eta, \Delta\}$ where
tan$\beta=v_2/v_1$.  (We have used eqs. (4) and (6) to eliminate $k$
in favor of the angle cos$\eta$.)

Since two of the neutral Higgs bosons $(h_1$ and $h_2$) are
relatively light in this model, their masses and mixings are
constrained by the LEP data on Higgs searches.  In fig.
2 is displayed a plot of the experimentally allowed region in the
$\lambda-{\rm cos}\eta$ plane for fixed values of the other parameters.
The first constraint is that $h_1$ and $h_2$ not have been produced in
the decay of a real $Z$ [10].  Since there is no significant suppression in
the $Zh_1h_2$ coupling, $Z \rightarrow h_1+h_2$ should be kinematically
forbidden.  The bottom (small $\lambda$) boundary of the allowed region
in fig. 2 corresponds to $m_{h_1}+m_{h_2} = M_Z$.  The second constraint is
that
a light Higgs particle not have been produced in the decay $Z
\rightarrow Z^* h$.  If $h=\sum_{i=1}^5\alpha_i\Phi_i$, where
$\sum_{i} |\alpha_i|^2 = 1$, then the cross section for the process is
approximately proportional to $|\alpha_1 {\rm sin}\beta+\alpha_2
{\rm cos}\beta|^2m_h^{-1}$.  The condition we have imposed, that gives
the upper (large $\lambda$) boundary of the allowed region is [10]
$m_h > (60~GeV) |\alpha_1 {\rm sin}\beta+\alpha_2{\rm cos}\beta|^2$.
In addition, the `pseudoscalar' should be heavier than 20 GeV [11],
which is always satisfied in the fit of fig. 2 (see Table 1).

The allowed region in fig. 2 corresponds to the fixed parameters having
the values tan$\beta=1, r=0.8, A=3v, x=3.8v$ and $\Delta = 3.8$.  This
is a relatively large value for $\Delta$, but is realized if the
top Yukawa is large, near its renormalization-group fixed point.
For example, if $M_{sq}=3~TeV$, $h_t = 1.05$
(at $Q^2 = (3~TeV)^2$)
and $p=1$ in eq. (10), we obtain $\Delta=3.8$ at $Q^2=m_t^2$.  (This
involves a further running from $Q^2=M_{sq}^2$ to $m_t^2$ assuming
non--SUSY two--doublet spectrum.)
A larger value of $h_t$ will be inconsistent with perturbative
unification (in the desert scenario), so a larger $\Delta$ seems to be
unlikely, unless the squark masses are heavier than $3~TeV$.  If
$\Delta$ is taken to be smaller (with the other parameters fixed at the
above values), the allowed region rapidly shrinks and vanishes at
$\Delta \simeq 3.1$ corresponding to $M_{sq} \simeq 1.5~TeV$ (with the same
$h_t$).
Similarly, if tan$\beta$ is increased above 1,
the allowed region shrinks and disappears at
tan$\beta\simeq 1.3$.  In this scenario of CP--violation, then, the
parameters $\Delta(M_{sq},h_t)$ and tan$\beta$ are tightly constrained.
The same is seen from fig. 2 to be true of the parameter $\lambda$
which must lie in a rather narrow range (around 0.2 for $r=0.8$).
Numerically it is found that as $r$ decreases the allowed range of
$\lambda$ shifts upward, and vice versa.  On the other hand, the
parameters $r,A,x$ and cos$\eta$ can vary over a rather wide range.  For
example, we find solutions for $0.5 \le r \le 2.7$.  Much of the above
behavior can be understood fairly well analytically.

Consider certain subdeterminants of the $5 \times 5$ mass--squared matrix
given in eq. (9).  If the CP--violating extremum is to be a local
minimum, this matrix must have no negative eigenvalues and its
subdeterminants must be all non--negative.
\begin{eqnarray}
{\rm det}_{1235} &=& 9 \lambda^2 r^2 A^2{\rm sin}^2 2\beta {\rm sin}^2\eta v^6
\left\{{1 \over 2}\lambda_1[{1 \over 2}\lambda_1\Delta-2(3r-1)
\lambda^2]-\lambda^4(3r-1)^2\right\} \\
{\rm det}_{345} &=& 3 \lambda^2r^3A^2x^{-2}(3r-1){\rm sin}^2\eta \{9
\lambda^2{\rm sin}^22\beta v^6 + 4 A^2x^2 v^2+
12 \lambda A {\rm sin}2\beta {\rm cos}\eta x v^4\} \nonumber \\
\end{eqnarray}
where det$_{ij..k}$ is the determinant of the submatrix corresponding to
the $ij..k^{\rm th}$ rows and columns.  Since $r \ge 0$ from
${\cal M}_{33}^2$, eq. (13) implies
that $(3r-1) \ge 0$.  If one sets $\Delta=0$, then the right hand side
of eq. (12) with $(3r-1) \ge 0$ is negative.  This is the result of
Ramao [6], that at tree--level spontaneous CP violation does not occur in
the model.

Eqs. (12)-(13) allow us to place bounds on $\lambda^2$ if CP is
spontaneously broken in the Higgs sector.
\begin{equation}
0 \le (3r-1)\lambda^2 \le {1 \over 2} \lambda_1\left(\sqrt{1+\Delta}-1
\right)~~.
\end{equation}
One also sees from eq. (12) and (13) that the light Higgs masses are
made largest by making sin$2\beta$ large, that is, tan$\beta \simeq 1$
is preferred.  As a consequence of eq. (14), the charged Higgs
boson mass is predicted to be less than about 110 GeV (see eq. 11).

More insight can be obtained by considering the small $\lambda$ limit,
which is justified by eq. (14) and by our numerical results.  As
$\lambda \rightarrow 0$, the mixing between the $\Phi_{1,2,3}$ block and
$\Phi_{4,5}$ block goes away.  If one block diagonalizes to obtain the
$3 \times 3$ block to order $\lambda^2$ one finds
\begin{eqnarray}
{\cal M}^2(\Phi_1,\Phi_2,\Phi_3) &=& \left(\matrix{\lambda_1v_1^2 &
-v_1v_2(\lambda_1+2(3r-1)\lambda^2) & 0 \cr
-v_1v_2(\lambda_1+2(3r-1)\lambda^2) & \lambda_1v_2^2(1+\Delta) & 0
\cr 0 & 0 & 0}\right)  \nonumber \\
& + & C \lambda_1 \left(\matrix{v_2^2{\rm cos}^2\eta & v_1v_2 {\rm cos}^2\eta &
-v v_2 {\rm sin}\eta {\rm cos} \eta \cr
v_1 v_2 {\rm cos}^2 \eta & v_1^2 {\rm cos}^2 \eta & -v v_1 {\rm sin}\eta
{\rm cos}\eta \cr -v v_2 {\rm sin} \eta {\rm cos}\eta & -v v_1
{\rm sin}\eta {\rm cos}\eta & v^2 {\rm sin}^2\eta}\right)
\end{eqnarray}
where $C \equiv \left[4r(3r-1)/(4r-1)\right]\left(\lambda^2/
\lambda_1\right)$.  From the
eigenvalues of ${\cal M}^2(\Phi_1,\Phi_2,\Phi_3)$  a
non--trivial lower bound on $\lambda^2$ can be derived as follows.
Label the eigenvalues
to be $M_I^2 \le M_{II}^2 \le M_{III}^2$.  It is possible
by considering the eigenvalues as a function of $C$ (defined above)
to show that $M_I^2 \le C M_Z^2 = C \lambda_1 v^2$.
(This can be done by plotting the eigenvalues of ${\cal M}^2(C)/M_Z^2$
versus $C$ and considering how often they cross the line $y=C$ which can
be determined from the equation det$\left\{{\cal M}^2(C)/M_Z^2-CI\right\}=0$.
This is easily seen to be quadratic (not cubic) in $C$ and to have two
real roots, one of which is $C=0$.)  From eq. (15) one sees that for
small $\lambda^2$ and $\Delta \ge {\rm cot}^2\beta-1$ the largest entry
of ${\cal M}^2$ is greater than $(1+\Delta)\lambda_1v_2^2 =
(1+\Delta)M_Z^2{\rm sin}^2\beta$.  Therefore $M_{III}^2 \ge (1+\Delta)M_Z^2
{\rm sin}^2\beta$.  The trace of ${\cal M}^2$ implies that $M_I^2
+M_{II}^2+M_{III}^2 = M_Z^2(1+\Delta{\rm sin}^2\beta + C)$.  Thus
$M_I^2+M_{II}^2 \le M_Z^2(C+{\rm cos}^2\beta)$.
This together with $M_I^2 \le CM_Z^2$
imply that if $C \le {\rm cos}^2\beta$, then
$(M_I+M_{II})/M_Z \le \sqrt{C}+{\rm cos}\beta$, while if $C \ge
{\rm cos}^2\beta$, then $(M_I+M_{II})/M_Z \le \sqrt{2C+2{\rm cos}^2\beta}
$.  In order to have $M_I+M_{II} \ge M_Z$ it then must be that
$C \ge (1-{\rm cos}\beta)^2$.  Using the definition of $C$ one gets a
lower bound on $\lambda^2$, which combining with eq. (14) gives
\begin{equation}
{{(4r-1)}\over{4r}}\lambda_1(1-{\rm cos}\beta)^2 \le (3r-1)\lambda^2 \le
{1 \over 2}\lambda_1(\sqrt{1+\Delta}-1)~.
\end{equation}
For the values of $r=0.8, \Delta=3.8, {\rm tan}\beta=1$ that are used
in fig. 2 one has $0.13 \le \lambda \le 0.41$ which is seen to be
satisfied well by the numerical results.  One can also derive from
eq. (16) a better lower limit on $r$ than $r \ge 1/3$.

A remark about the small $\lambda$ limit is in order.  From eq. (4) and
fig.1 it can be seen that as $\lambda \rightarrow 0$, $x/A \rightarrow
r/k$.  ($k$ in fig. 2 is about $0.63$, the fixed point value.)  Terms
have been dropped in deriving eq. (15) that are down by
$O(\lambda x/A)$.  For large $r$, these terms become important, which
tend to decrease the light eigenvalues of ${\cal M}^2$.  This is why no
realistic solution exists for large
$r$ ($r \stackrel{_>}{_\sim} 2.7$).

In fig. 2 we have also plotted contours of the electric dipole moment of
the electron.  These have been computed using the results of Leigh,
Paban and Xu (LPX) [1].  Only the graphs with the same topology as the top
quark graphs (figs. 1 and 2 of LPX), and only the
contributions from the lightest two neutral Higgs have been included.
This should be good to about 20\%.  The typical values of $d_e$ are in
the range $(10^{-28}~{\rm to}~3 \times 10^{-27})$ e-cm, interesting for
future atomic experiments.  In Table 1 are presented the masses and
mixings of the lightest Higgs for various $\lambda$ and cos$\eta$ in fig.
2 from which we have computed $d_e$.

It is interesting to ask if the Higgs sector CP--violation alone
can account for $\epsilon$ and $\epsilon^{\prime}$ in the K meson
system.  Pomarol [3] has shown that this may indeed be possible in a
more generalized version of the SUSY singlet model.
Since the effects of spontaneous CP violation felt in the fermion
sector in our model is identical, we expect the results of Ref. 3
to hold in our case as well.

As for the neutron edm ($d_n$), although it is difficult to predict its
value precisely due to strong interaction uncertainties, we expect $d_n$
to be near the experimental limit, $d_n \sim 10^{-26}$ e-cm.  Indeed, if
the only source of CP--violation is in the Higgs sector, then the usual
fine--tuning of the phase of the gluino mass term can be avoided.  This
phase will be zero at tree--level, and will arise only at one--loop,
giving experimentally consistent value for $d_n$.

So far we have said nothing about whether the CP--violating minimum that
we have been studying is also the absolute minimum.  For fixed $|H_1|,
|H_2|$ and $|N|$ it is easy to see that $V$ at the CP--violating extremal
angles (eq. 6) is lower than at the CP conserving values $\eta, \zeta = 0,
\pi$.  Of
course the true CP--violating and CP--conserving minima will not be at
the same values of $H_1,H_2$ and $N$. However, it can be shown
analytically that for small $\lambda$ the CP--violating minimum is the
lowest [9].  We have verified this result numerically for some reasonable
values of the parameters.

\section*{References}
\begin{enumerate}
\item S. Weinberg, Phys. Rev. Lett. {\bf 63}, 2333 (1989);
S.M. Barr and A. Zee, Phys. Rev. Lett. {\bf 65}, 21 (1990); J.
Gunion and R. Vega, Phys. Lett. {\bf 251B}, 157 (1990); D. Chang, W.-Y.
Keung and T.C. Yuan, Phys. Rev. {\bf D43}, 14 (1991); R.G. Leigh, S.
Paban and R-M. Xu, Nucl. Phys. {\bf B352}, 45 (1991).  For a review of
atomic searches, see S.M. Barr, Int. J. Mod. Phys. {\bf A8}, 209
(1993).
\item V.A. Kuzmin, V.A. Rubakov and M.E. Shaposhnikov, Phys. Lett.
{\bf 155B}, 36 (1985); M.E. Shaposhnikov, Nucl. Phys. {\bf B287}, 757
(1987); {\bf B299}, 797 (1988); P. Arnold and L. McLerran, Phys. Rev.
{\bf D 36}, 581 (1987); {\bf D 37}, 1020 (1988); N. Turok and J.
Zadronzny, Phys. Rev. Lett. {\bf 65}, 2331 (1990); A.G. Cohen, D.B.
Kaplan and A.E. Nelson, Phys. Lett. {\bf 263B}, 86 (1991).
\item See for example, A. Pomarol, Phys. Rev. {\bf D 47}, 273 (1993).
\item It has been argued that
spontaneous CP violation may occur in the MSSM after radiative
corrections are included, see N. Maekawa, Phys. Lett. {\bf B282}, 387 (1992).
However,  this is inconsistent with LEP
limits on light Higgs searches,
see A. Pomarol, Phys. Lett. {\bf B287}, 331 (1992).  See also G. Branco
and N. Oshimo, Lisbon Preprint IFM 1/93 (1993).
\item H. Nilles, M. Srednicki and D. Wyler, Phys. Lett. {\bf B120}, 346
(1983); J.P. Derendinger and C. Savoy, Nucl Phys. {\bf B237}, 307
(1984); J. Gunion and H. Haber, Nucl. Phys.
{\bf B272}, 1 (1986);
J. Ellis, J. Gunion, H. Haber, L. Roskowski and F. Zwirner,
Phys. Rev. {\bf D39}, 844 (1989);
M. Drees, Int. J. Mod. Phys.
{\bf A4}, 3635 (1989); L. Durand and J. Lopez, Phys. Lett. {\bf B217},
463 (1989); J. Espinosa and M. Quiros, $ibid$. {\bf B279}, 92
(1992); T. Moroi and Y. Okada, $ibid$. {\bf B295}, 73 (1992);
U. Ellwanger and M. Rausch de Traubenberg, Z. Phys.
{\bf C53}, 521 (1992); T. Elliot, S.F. King and P.L. White, Phys. Lett.
{\bf B305}, 71 (1993);
G. Kane, C. Kolda and J. Wells, Phys. Rev. Lett.
{\bf 70}, 2686 (1993); W. ter Veldhuis, Purdue Preprint PURD-TH-92-11;
C. Savoy, SACLAY Preprint, T93/060 (1993).
\item  J. Ramao, Phys. Lett. {\bf B173}, 309 (1986).
\item Superficially this would seem to contradict the Georgi-Pais
theorem: H. Georgi and A. Pais, Phys. Rev. {\bf D 10}, 1246 (1974);
{\bf D 16}, 3520 (1977).  But it does not since the radiative
corrections here are not small in the sense assumed in the proof of that
theorem.
\item In supergravity models $r=1$ (and thus real) at the unification
scale.  $r$ can
develop a phase in the process of running only through
complex gaugino mass terms, but this effect will be small due to
constraints on the gaugino phase from the neutron EDM.
\item Details can be found in a forthcoming paper, K.S. Babu and S.M.
Barr, in preparation.
\item K. Hikasa et. al., (Particle Data Group), Phys. Rev. {\bf D 45},
S1 (1992).
\item D. Decamp et.al., (ALEPH Collaboration), Phys. Lett. {\bf B265},
475 (1991).
\end{enumerate}

\newpage
\section*{Figure Captions}

\hspace*{0.2in} Fig. 1.  This triangle shows the relation between
the extremal values of
the phase angles $\eta$ and $\zeta$ and the magnitudes of the fields and
the parameters of the Higgs potential.  If $D,E$ and $F$ are not in such
a relation that this triangle can be drawn, then there is no
CP--violating extremum.

Fig. 2.  A plot of the allowed region (inside the closed curves) of the
$(\lambda-{\rm cos}\eta$) plane, for tan$\beta=1, r=0.8, A=3v, x=3.8v$
and $\Delta=3.8$.  The lower boundary of the region corresponds to
$m_{h_1}+m_{h_2}=M_Z$.  The upper boundary is the constraint from
non--observation of $Z \rightarrow Z^* +h$.  The contours inside the
region represent the electron EDM in units of $10^{-27}$ e-cm.
\vspace{.14in}

\noindent{\large \bf Table 1}.  Some selected
points in the ($\lambda - {\rm cos}\eta$) plane
for the same parameter values as in fig. 2.  $m_{h_1}$ and $m_{h_2}$ are the
masses (in GeV) of the lightest two neutral Higgs $h_1$ and $h_2$.  The
$R_{ij}$ are defined by $h_i = \sum_{i=1}^5 R_{ij}\Phi_j$.  The electron
EDM $d_e$ is in units of $10^{-27}$e-cm.
\vspace{.14in}

$$\begin{array}{|c|c|c|c|c|c|c|c|c|c|c|}\hline
{\rm cos}\eta & \lambda & m_{h_1} & m_{h_2} & R_{11} & R_{12}
& R_{13} & R_{21} & R_{22} & R_{23} & d_e \\\hline
-0.5 & 0.235 & 34.3 & 56.9 & -0.78 & -0.29 & 0.55 & 0.53 & 0.17 & 0.83 &
-10.0 \\
0.0 & 0.17 & 49.5 & 41.6 & -0.92 & -0.29 & 0.25 & 0.24 & 0.08 & 0.97 &
1.75 \\
0.0 & 0.21 & 42.5 & 53.3 & 0.88 & 0.31 & 0.35 & -0.33 & -0.11 & 0.94 & 3.0
\\
0.5 & 0.195 & 59.2 & 32.0 & -0.82 & -0.24 & 0.52 & 0.49 & 0.18 & 0.85 & 11.5
\\
0.5 & 0.225 & 63.2 & 30.6 & -0.73 & -0.21 & 0.65 & 0.61 & 0.23 & 0.76 & 15.2
\\
0.5 & 0.253 & 68.5 & 26.2 & -0.67 & -0.19 & 0.74 & 0.68 & 0.27 & 0.68 & 21.5
\\
0.67 & 0.223 & 66.7 & 24.5 & -0.81 & -0.23 & 0.55 & 0.51 & 0.20 & 0.84 & 19.8
\\ \hline
\end{array}
$$

\end{document}